# Comparative Benchmarking of Utility-Scale Quantum Emulators

ANNA LEONTEVA, GUIDO MASELLA, and MAXIME OUTTERYCK, QPerfect SAS, France
ASIER PIÑEIRO ORIOLI and SHANNON WHITLOCK, European Center for Quantum Sciences
(CESQ-ISIS, UMR 7006), University of Strasbourg and CNRS, and QPerfect SAS, France

Evaluating quantum algorithms at utility-scale – involving more than 100 qubits – is a key step toward advancing real-world applications of quantum computing. In this study, we benchmark seven state-of-the-art quantum emulators employing techniques such as tensor networks, matrix product states (MPS), decision diagrams, and factorized ket based methods, running on CPU based hardware and focusing on effectively exact simulations. Performance is assessed on 13 benchmark circuits from the MQTBench library, spanning circuit sizes from 4 to 1,024 qubits. Our results reveal that MPS-based emulators outperform other approaches overall, successfully solving 8 benchmarks up to the maximum size of 1,024 qubits and 12 benchmarks up to at least 100 qubits in less than 5 minutes. We find evidence that all circuits except a random one can be simulated in polynomial time. This work demonstrates that quantum emulators can faithfully simulate a broad range of large and complex universal quantum circuits with high fidelity, far beyond the limits of statevector simulators and today's quantum hardware.

## 1 INTRODUCTION

### 1.1 Context and Importance of Quantum Emulation

Quantum computing promises to unlock solutions to problems that are intractable for classical computers [1], potentially revolutionizing fields such as scientific research [2–5], finance [6, 7], and machine learning [8–10]. However, current quantum hardware is in its early stages, limited by high gate error rates and significant challenges in scaling to large numbers of qubits. In the meantime, quantum emulators – software designed to simulate quantum computers on classical hardware – offer a valuable tool for developing and testing quantum algorithms prior to their deployment on actual quantum processors.

The use of classical computers to simulate quantum systems has a long history, predating Feynman's seminal lecture "Simulating Physics with Computers" [11], which laid the foundation for quantum computing by highlighting the inefficiency of classical methods for simulating quantum dynamics. Over the intervening decades, various numerical methods, such as exact diagonalization, Monte Carlo methods, density matrix renormalisation group algorithms, and Tensor Network (TN) representations such as Matrix Product States (MPS) [12–16], have been developed to model quantum systems at larger and larger scales, aided by advances in High-Performance Computing (HPC). Today, classical simulation plays a pivotal role in quantum computing, providing a practical testing ground for refining quantum algorithms, understanding and mitigating the impact of noise, and benchmarking quantum processors at the limits of classical computing [17–21].

Utility-scale quantum computing [18] refers to the capability of quantum processors to address large-scale, real-world computational problems that surpass the performance of classical methods. Typically, this requires deep quantum circuits with 100 or more qubits [22]. However, the boundary is not as clear cut, as ongoing advances in numerical methods for simulating quantum circuits continue to push the limits [23–26]. In some cases these methods are optimized for specific classes of problems, while others promise broad applicability across a wide range of problems.

Previous benchmarking studies have addressed different aspects of quantum emulation on small scales, largely focusing on statevector methods and narrow application-specific problems. For example, Wu et al. [27] benchmarked statevector emulators up to 30 qubits, while Vallero et al. [28] evaluated GPU-accelerated quantum subroutines, including TN emulators, but restricted to problems with fewer than 30 qubits. Similarly, Jamadagni and Läuchli evaluated various simulation software packages up to 34 qubits on a small HPC cluster [29]. In parallel, a number of projects have put forward benchmarking toolkits and libraries for assessing quantum computing applications. Finžgar et al. [30] proposed an application-driven benchmarking method targeting two industrial problems. This was later extended by Kiwit et al. [31] to include quantum machine learning algorithms with GPU-accelerated statevector emulators. Lubinski et al. [32] further developed an application-oriented benchmarking methodology by evaluating a wide range of quantum algorithms applied to IBM's emulator and Quantinuum H1-1 hardware with circuit sizes up to 14 qubits. However, until now, there has been no similarly comprehensive study of quantum emulators at the utility-scale.

### 1.2 Objectives and Methodology

This study presents the first comprehensive evaluation of quantum emulators across a diverse set of scalable quantum circuits at the utility-scale (100+ qubits), establishing practical performance boundaries for these methods on commodity hardware. Our work addresses three core objectives:

- Develop a transparent, user-centric benchmarking methodology for quantum emulators (which could even be extended to quantum computers).
- Systematically evaluate and compare emulator performance on 100+ qubit circuits using three metrics: total run time, accuracy, and scalability.
- Classify which families of quantum algorithms can be simulated efficiently on classical computers versus those requiring genuine quantum hardware.

To achieve these objectives we selected seven utility-scale emulators from six leading frameworks focusing on the most relevant numerical methods, applied across 13 circuit classes scalable from 4 to 1,024 qubits.

## 2 BENCHMARK METHODOLOGY

### 2.1 Selected Emulators

Our study covers a set of seven actively supported quantum emulators that extend beyond the statevector approach: four Matrix Product State (MPS) emulators (`Qiskit-MPS`, `Quimb-MPS`, `QMatchaTea-MPS`, as well as `MIMIQ-MPS`, a commercial emulator developed by the authors); one Tensor Network (TN) emulator (`Quimb-TN`); one factorized ket based emulator (`Pyqrack`); and one Decision Diagrams (DD) emulator (`MQT-DDS`). These emulators were selected to cover multiple approaches and on the basis that they were capable of scaling to beyond 100 qubits within 5 minutes, for at least some of the tested circuits, thereby ensuring their relevance for utility-scale problems. A detailed list of the selected utility-scale emulators, as well as other emulators which were considered, but not included in the study, is provided in Appendix A.

### 2.2 Benchmarking Suite

As a source of problem sets, we selected the open-source `MQTBench` library[1] [33], a standardized framework designed to assess the performance of quantum computing platforms by providing scalable quantum circuits in `OpenQASM 2.0` [34] format. The suite covers a range of commonly used quantum circuit primitives and application-oriented tasks, from which we identify 13 circuit classes capable of scaling beyond 100 qubits. To push the computational limits of the selected emulators, we extended the circuit sizes beyond the 130-qubit constraint of the default `MQTBench` suite using the library's published source code. To ensure compatibility across all selected emulators, each algorithm was transpiled to a minimal gate set (comprising U and CX gates) and sanitized by removing gates with negligible rotation angles and normalizing large angles modulo $4\pi$. A detailed list of the circuits in the benchmark suite is provided in Appendix B and all QASM files used in this study along with their corresponding mirror circuits are available on Zenodo[2].

---
[1]https://github.com/cda-tum/MQTBench

### 2.3 Performance Metrics

*2.3.1 Scalability.* Scalability in quantum simulation quantifies an emulator's ability to efficiently handle increasing problem sizes without loss of accuracy or excessive run time. To measure scalability, we evaluate how the run time depends on the number of qubits $n$ for each problem set, subject to the constraints that the overall fidelity of the simulation is at least 0.99 and the maximum run time does not exceed 300 seconds:

$$\mathcal{T}_{\mathcal{E}}(n) = \text{Run time of emulator } \mathcal{E} \text{ for } n \text{ qubits, subject to } \mathcal{F}_{\mathcal{E}}(n) \geq 0.99, \qquad (1)$$

where $\mathcal{F}_{\mathcal{E}}(n)$ is the overall circuit fidelity achieved by emulator $\mathcal{E}$ when executing an $n$-qubit circuit. Emulators that successfully simulate circuits with run times scaling polynomially with $n$ (appearing as a straight line on a log-log run time plot) are considered more scalable than those with exponential scaling. Additionally, we evaluate the maximum number of qubits successfully simulated by each emulator and each circuit class, up to the largest benchmark circuit size of 1,024 qubits.

*2.3.2 Run Time.* For the purposes of this study, run time is defined as the total wall-clock time required to execute a quantum circuit from initialization to final measurement (excluding mirror circuit fidelity estimation, below). This encompasses all necessary pre-processing steps, including circuit parsing, data loading, circuit pre-optimization, state initialization, execution and sampling. Each benchmark is repeated four times, and the minimum run time (in seconds) is reported to account for warm-up and shot-to-shot variability.

*2.3.3 Mirror-Circuit Fidelity.* Mirror-circuit fidelity [35] provides a method to evaluate the accuracy of quantum emulators without relying on internal fidelity estimates, by executing a circuit in both forward and reverse directions. The probability of sampling the initial state yields the mirror-circuit fidelity – an independent accuracy test applicable across all emulators and benchmark problems, measuring how well the simulation preserves quantum information while minimizing significant errors. For each emulator we include barrier operations after the forward circuit (where supported), to ensure that an emulator does not simply compile the mirror circuit to an identity operation, and check that the emulator produces viable samples consistent with the other emulators after the forward circuit.

### 2.4 Automated Hyperparameter Optimization

Unlike statevector emulators which typically have fixed configurations, many of the selected emulators expose tunable parameters such as bond dimension, accuracy cutoffs, circuit pre-optimization and other method specific parameters. Their performance can be highly sensitive to these parameters. Therefore, to find good parameter sets without introducing bias, we integrate an automated optimization process using the Covariance Matrix Adaptation Evolution Strategy (CMA-ES) [36], which is well suited for noisy cost functions like run time. We use CMA-ES to minimize $\mathcal{T}_{\mathcal{E}}$ for a large problem instance while enforcing the fidelity constraint. While CMA-ES is originally designed for continuous optimization problems, we adapt it to discrete parameters by mapping each continuous candidate solution to the nearest valid discrete configuration and applying a margin correction to adjust the covariance matrix [37] for preventing premature convergence. The parameters found by automatic optimization are similar to what we found via a manual search, and yield better or at least comparable results. The optimizer is made available as part of our open-source repository on GitHub.

---
[2]https://doi.org/10.5281/zenodo.15220683

## 2.5 Benchmarking Protocol

The finalized benchmarking protocol consists of the following steps:

(1) Hyper-parameter optimization. We employ the `CMA-ES` evolutionary strategy to optimize emulator hyperparameters on a 100 qubit problem instance, minimizing execution time while maintaining a minimum fidelity threshold of $\mathcal{F} \geq 0.99$. If the run time exceeds 300 seconds, we fall back to a problem size of 24 qubits for optimization (10 qubits for the `random` circuit). We find that for the vast majority of problems, the 100-qubit optimized hyperparameters work well, even for larger instances, as verified by achieving a fidelity $\mathcal{F} \geq 0.99$. Cases where this goal is not met, as well as other early termination conditions are documented in Appendix F.
(2) Execution. Each of the 13 circuit classes contains multiple circuit instances defined for different number of qubits. Problem sizes range from 4 to the maximum size of 1,024 qubits and are sampled on a quasi-logarithmic grid to reduce the total number of circuits benchmarked. A single execution of a circuit instance consists of a complete quantum computing workflow, which includes parsing the `QASM` file, applying any available pre-optimization (if supported by the emulator), running the circuit, and sampling 1,000 measurement outcomes. For each execution we measure the total wall-clock time, including the time spent on `QASM` parsing, pre-optimization, circuit execution, and generating the samples. To reduce the impact of system variability we execute each circuit instance 4 times and take the minimum wall-clock run time achieved. Each circuit execution is subject to a strict 300-second timeout. Emulators exceeding this limit are classified as incapable of handling the target scale.
(3) Fidelity estimation via mirror circuits. We employ mirror circuits to obtain the circuit fidelity by executing both forward and reverse circuit iterations. Circuits failing to achieve $\mathcal{F} \geq 0.99$ return probability to the initial state are flagged as inaccurate at the tested scale.
(4) Data storage and analysis. All metrics – including run time and fidelity for each execution – are stored in structured JSON format, and available on Zenodo[3], for reproducible analysis and post-processing.

## 2.6 Hardware Setup:

All experiments were conducted on the same CPU-based system: an AMD EPYC 4244P 6-core processor with 12 threads, 130 GB RAM, running AlmaLinux 9.5. For all experiments, the number of threads was set to the maximum (12).

## 3 RESULTS

### 3.1 Pre-Processing: Optimization of Tuning Parameters

Table 1 summarizes the results of optimizing the emulator hyperparameters using `CMA-ES` across all 13 benchmark circuits. For non-MPS-based emulators, `Quimb-TN` required optimization of its single Boolean hyperparameter ("`Contract`"), which we did manually, while `MQT-DDS` and `Pyqrack` had no tunable parameters. The optimal hyperparameters for each emulator vary widely across circuit classes. This underscores the importance of good hyperparameter optimization, or in its absence, knowledge of numerical methods and the structure of the circuit being simulated, to obtain the best

---
[3]https://doi.org/10.5281/zenodo.15233671

Comparative Benchmarking of Utility-Scale Quantum Emulators            5Table 1. **Optimized tuning parameters for MPS-based and tensor network emulators.** In each cell, the first bracket lists the convergence parameters and the second lists the optimization parameters. For `MIMIQ-MPS`, the format is [bond_dimension / entdim / scut], [meth / fuse / perm]. For `Qiskit-MPS`, the format is [matrix_product_state_max_bond_dimension / matrix_product_state_truncation_threshold / mps_sample_measure_algorithm], [ optimization_level / mps_lapack]. For `Quimb-MPS`, the format is [bonddim / cutoff], [permute / gate_contract]. For `QMatchaTea-MPS`, the format is [max_bond_dimension / cut_ratio], [optimization_level / tensor_compilator_enable / linearize_enable]. For, `Quimb-TN`, a single parameter, [Contract].

| **Circuit** | `MIMIQ-MPS` | `Qiskit-MPS` | `Quimb-MPS` | `QMatchaTea-MPS` | `Quimb-TN` |
|---|---|---|---|---|---|
| qft | [4/4/1e-5], [vmpoa/T/F] | [4/1e-10/AM],[1/T] | [4/1e-10], [F/AU] | [4/1e-10], [1/T/T] | F |
| qftentangled | [4/4/1e-5],[vmpoa/T/F] | [4/1e-10/AM],[1/F] | [4/1e-10], [F/AU] | [4/1e-10],[1/T/T] | T |
| ghz | [4/4/1e-5], [vmpoa/T/F] | [4/1e-10/P],[0/T] | [4/1e-10],[F/AU] | [4/1e-10], [0/F/F] | F |
| wstate | [4/4/1e-5], [vmpoa/T/F] | [4/1e-10/P],[0/T] | [4/1e-10] [F/AU] | [4/1e-10], [0/F/F] | F |
| qpeexact | [4/4/1e-5],[vmpoa/T/F] | [4/1e-10/P],[1/T] | [4/1e-10],[F/AU] | [4/1e-10],[1/T/T] | F |
| qpeinexact | [4/4/1e-5],[vmpoa/T/F] | [4/1e-10/P],[1/T] | [8/1e-10],[F/AU] | [4/1e-10], [1/T/T] | F |
| qwalk | [32/8/1e-5],[vmpoa/T/F] | [32/1e-5/P],[0/F] | [8/1e-10],[F/SS] | [8/1e-10], [0/T/T] | T |
| ae | [64/4/1e-3],[vmpoa/T/F] | [32/1e-5/P],[1/F] | [32/1e-6],[F/AU] | [32/1e-2],[1/T/T] | F |
| realamp | [32/4/1e-5],[vmpoa/T/F] | [1024/1e-6/P],[1/T] | [32/1e-10],[F/AU] | [64/1e-10],[1/T/T] | T |
| su2rand | [32/4/1e-5],[vmpoa/T/F] | [1024/1e-6/P],[1/T] | [32/1e-10],[F/AU] | [64/1e-10],[1/T/T] | T |
| qnn | [384/4/1e-5], [dmpo/T/F] | [256/1e-5/P],[1/T] | [256/1e-10],[F/AU] | [256/1e-10],[2/T/T] | T |
| graphstate | [256/16/1e-5],[vmpoa/T/T] | [2048/1e-10/P],[1/T] | [512/1e-10],[T/AU] | [1024/1e-10],[1/F/T] | F |
| random | [512/8/1e-5],[vmpoa/T/T] | [2048/1e-5/P],[1/T] | [128/1e-10],[F/AU] | [1024/1e-10],[0/T/T] | T |

results from a given emulator. Based on our experiments, `CMA-ES` typically converges to an optimal solution within 4–5 iterations with a population size of 10, reflecting the low dimensionality of the tuning space. We also observed that multiple parameter combinations yield similar performance, suggesting that some parameters have limited influence while others are more critical.

We broadly classify hyperparameters according to two main groups: (i) convergence parameters: common settings, such as bond dimension and cut-off parameters; (ii) method parameters: emulator-specific configurations and circuit pre-optimization parameters that further enhance performance. A more detailed description of these parameters is provided in Appendix C. For MPS-based emulators, optimal values for the convergence parameters vary considerably between circuit classes, but are similar across emulators (with some exceptions). This indicates, for example, that bond dimension may be considered as an indicator for hardness. The easiest circuits are `qft` circuits, `ghz` and `wstate`, which generate low to moderate entanglement. We found `qpeexact` and `qpeinexact` are also exactly simulable with surprisingly small bond dimension, which we attribute to the specific benchmark which involves a single qubit in the state register and $n-1$ counting qubits. The circuits which require the highest bond dimensions overall are `qnn`, `random`, and `graphstate` suggesting that these circuits generate a lot of entanglement. But they also exhibit a lot of variability in optimal bond dimension which suggests circuit pre-optimization or different MPS methods may have a big impact on hardness. Appendix D provides a rough classification of the different circuits according to their simulation difficulty.

## 3.2 Run time Benchmark Results

We evaluated seven quantum emulators across 13 circuit classes to assess their minimum run time $\mathcal{T}_\mathcal{E}$ as a function of qubit count, subject to achieving a mirror circuit fidelity exceeding $\mathcal{F} \geq 0.99$. Figure 1 presents exemplary data for four circuit classes of varying difficulty: an easy task (`wstate`), a medium task (`realamp`), a hard task (`qnn`), and a very hard task (`random`). The shape of each curve provides valuable insights into emulator performance and the inherent classical difficulty of different quantum algorithms.

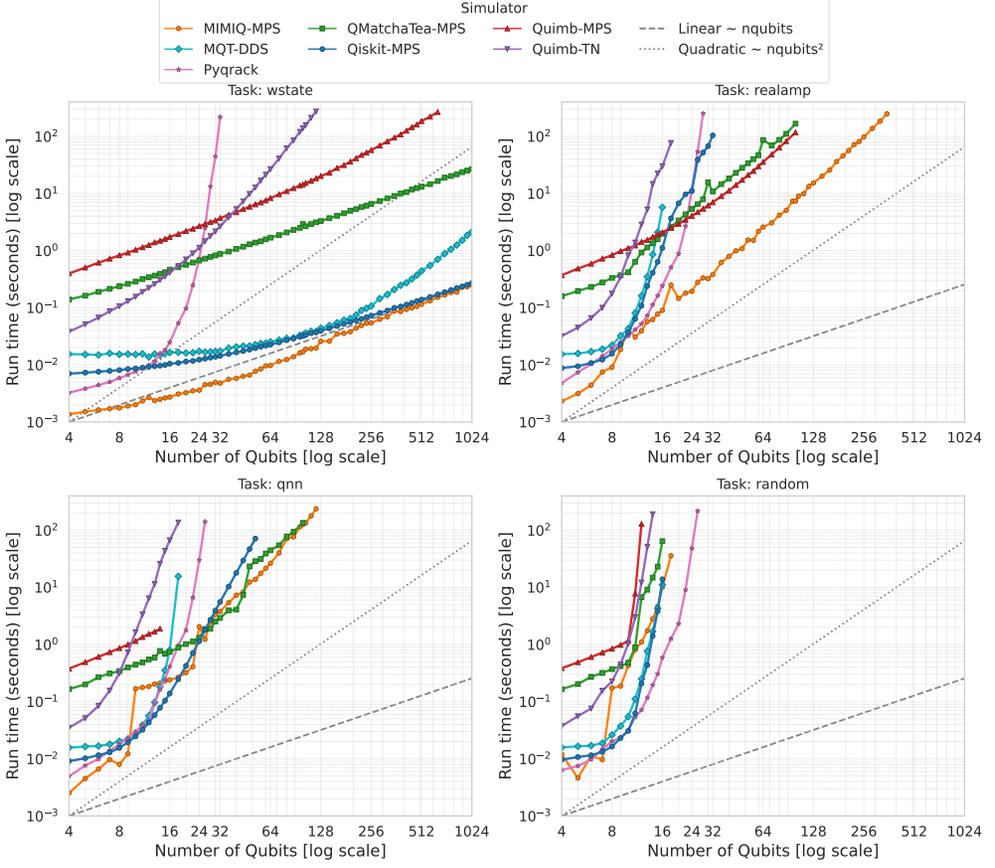

Fig. 1. **Run time plots comparing emulators as a function of qubit count on four benchmark circuits.** Each sub-panel corresponds to circuit classes with different difficulty levels: wstate (easy, top-left panel), realamp (medium, top-right panel), qnn (hard, bottom-left panel), and random (very hard, bottom-right panel). Each panel shows run time in seconds (log scale) versus the number of qubits (log scale) for the seven evaluated emulators indicated in the top legend. The dashed and dotted lines indicate linear and quadratic scaling laws as reference.

Generally we observe two different types of behavior for high and low qubit counts, with a crossover around 20 qubits. For qubit counts below 20, run time is presumably constrained by method-specific overheads and well-optimized statevector emulators likely outperform more sophisticated methods. Consequently, the following analysis focuses on problems with more than 30 qubits, where asymptotic scaling behavior starts to become apparent.

For the easy task (wstate), the majority of emulators exhibit polynomial scaling, evident as straight lines on a log-log plot. Notably, Matrix Product State (MPS) based emulators such as MIMIQ-MPS, Qiskit-MPS, and QMatchaTea-MPS demonstrate near-linear scaling ($T \propto n$), albeit with varying pre-factors. This is remarkable considering that both the gate count and circuit depth for this algorithm also scale linearly with $n$ implying the time to apply each gate is effectively independent of $n$. In contrast, Quimb-MPS and non-MPS emulators show superlinear scaling for this circuit.

Table 2. **Run time of all emulators across 13 benchmark algorithms at a scale of 100 qubits in seconds.** The best-performing emulator for each problem set is highlighted in bold font.

| Algorithm | MQT-DDS | Pyqrack | Qiskit-MPS | MIMIQ-MPS | Quimb-MPS | QMatchaTea-MPS | Quimb-TN |
|---|---|---|---|---|---|---|---|
| qwalk | **1.78** | – | 4.00 | 20.36 | – | – | – |
| qnn | – | – | – | **126.24** | – | 134.59 | – |
| realamp | – | – | – | **7.37** | 117.53 | 167.40 | – |
| su2rand | – | – | – | **8.26** | 119.89 | 174.15 | – |
| ae | – | – | 0.89 | **0.87** | 61.54 | 12.69 | – |
| qftentangled | – | – | **0.40** | 0.84 | 62.06 | 11.94 | – |
| qpeexact | – | – | **0.34** | 0.78 | 62.4 | 18.97 | – |
| qpeinexact | – | – | **0.44** | 0.82 | 63.27 | 18.94 | – |
| qft | – | – | **0.40** | 0.71 | 61.60 | 12.13 | – |
| graphstate | – | – | – | **0.25** | – | – | 175.20 |
| wstate | 0.03 | – | 0.03 | **0.01** | 14.39 | 2.92 | 138.50 |
| ghz | 0.02 | 5.94 | 0.02 | **0.003** | 14.34 | 2.82 | 2.58 |
| random | – | – | – | – | – | – | – |

The medium task (realamp) is particularly intriguing. This circuit, frequently employed as an ansatz in quantum chemistry and machine learning applications, comprises random rotations and entangling gates with all-to-all qubit connectivity. A priori, we don't expect this to be easy to classically simulate, however three emulators successfully simulate this circuit with polynomial run time scaling with respect to qubit number (approximately quadratic). This task also shows the most significant performance disparity among emulators, with nearly three orders of magnitude difference between Qiskit-MPS and MIMIQ-MPS at 32 qubits. This underscores how emulators can yield dramatically different performance outcomes even if they are based on the same underlying methods.

The hard task (qnn) displays qualitatively similar behavior to realamp. However only two emulators, QMatchaTea-MPS and MIMIQ-MPS, successfully simulate 100 qubits within the 300-second run time limit, with a power-law exponent closer to 3. The hardest task (random) proves challenging for all emulators, each exhibiting exponential scaling in qubit number. The best performance is achieved by Pyqrack, reaching 26 qubits in under 300 seconds.

Table 2 presents an overview of all benchmark problems evaluated at the 100 qubit scale. Each entry is the run time (in seconds) required to successfully execute a circuit with fidelity $\mathcal{F} \geq 0.99$. No value is given in cases where the emulator was unable to successfully execute the 100 qubit instance in under 300 seconds. Detailed run time results for all circuits for different qubit numbers are provided in Appendix H.

Overall we find that MPS-based emulators exhibit polynomial scaling, typically with a low power-law exponent, for 12 out of 13 algorithms (for at least one MPS emulator). This finding is unexpected, suggesting that many benchmarks in the MQTBench library admit efficient classical implementations. The non-MPS emulator Quimb-TN also exhibits polynomial scaling for several benchmarks, albeit with a large exponent. In general, the non-MPS emulators demonstrate exponential scaling, except for a few easy circuits such as ghz, wstate and qft, and MQT-DDS for qwalk.

The benchmarking results reveal substantial variability in performance among quantum emulators across different circuit types. MIMIQ-MPS demonstrated the broadest applicability, successfully solving 12 benchmark problems at the 100 qubit scale, followed by QMatchaTea-MPS and Quimb-MPS, which solved 10 and 9 problems respectively, and Qiskit-MPS, which solved 8 problems. In terms of run times, `MIMIQ-MPS` and `Qiskit-MPS` consistently outperformed other MPS-based emulators, typically 20-80 times faster on comparable problems.

Among non-MPS-based methods, `MQT-DDS` and `Quimb-TN` each solved three benchmark problems at the 100-qubit scale, while Pyqrack successfully solved one problem within the run time limit of 300 seconds. Despite the strong performance of MPS-based emulators overall, it is noteworthy that no single emulator consistently dominated in all circuit classes. This suggests that the limits of emulator performance have not yet been reached, at least for general quantum circuits, highlighting opportunities for future development in quantum emulator design.

## 3.3 Scalability Benchmark Results

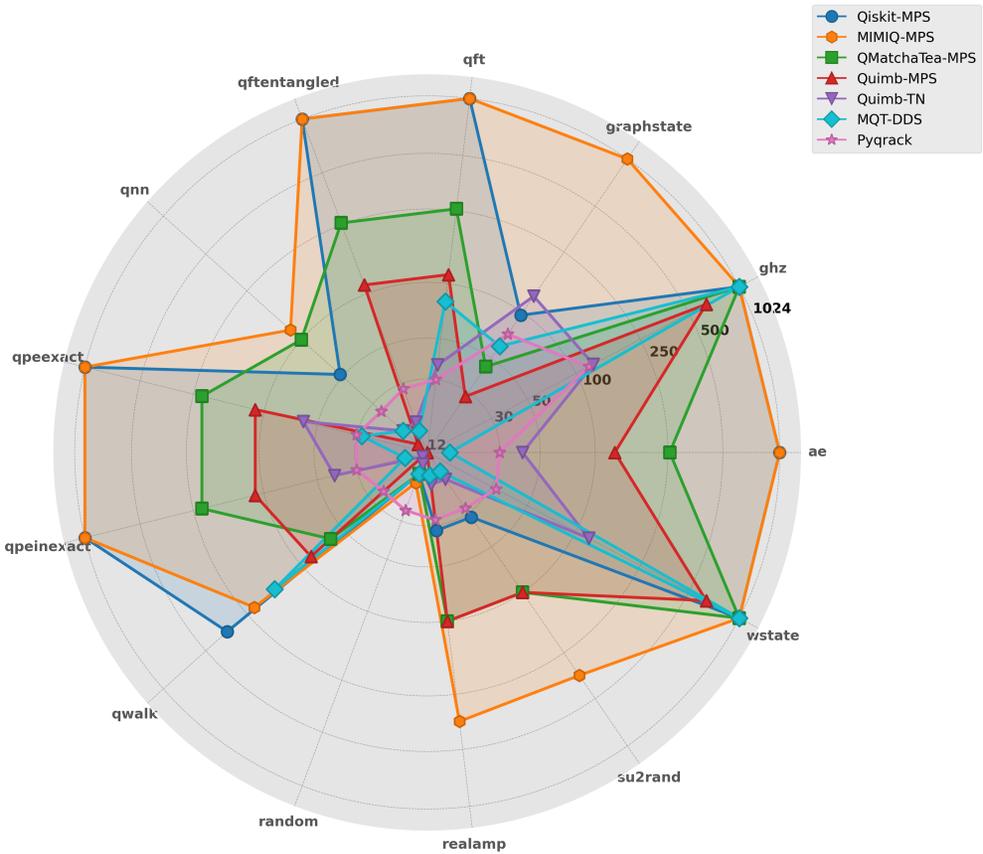

Fig. 2. **Emulators scalability across algorithms**. Radar chart comparing the maximum achievable qubit counts for each emulator across the 13 benchmark circuits. Each radial axis represents a different benchmark algorithm, with concentric circles indicating qubit scale from 12 to 1,024 qubits on a logarithmic scale. Different colored polygons represent different emulators: `Qiskit-MPS` (blue circles), `MIMIQ-MPS` (orange hexagons), `QMatchaTea-MPS` (green squares), `Quimb-MPS` (red triangles), `Quimb-TN` (purple downward-pointing triangles), `MQT-DDS` (light blue diamonds), and `Pyqrack` (pink stars). Larger polygon areas indicate better overall scalability.

To conclude our analysis, Figure 2 illustrates the scalability of each emulator across all 13 benchmark problem sets. The radial coordinate depicts the maximum qubit count achieved on a logarithmic scale with the innermost circuit corresponding to 12 qubits and the outermost circle corresponding to 1,024 qubits. This visualization provides an immediate overview of the fraction of benchmark circuits successfully solved by each emulator, with larger polygon areas representing more scalable emulators. Detailed data on maximum qubit counts for each emulator and circuit class are provided in Table 4 in Appendix E. In terms of scalability, Matrix Product State (MPS) methods demonstrate the broadest applicability. For instance, `MIMIQ-MPS` successfully solves the largest problem instances of 1,024 qubits for 8 out of 13 circuit classes, and likely more if the 300-second runtime constraint is relaxed. Non-MPS methods achieve this scale on only a limited number of circuits, notably `MQT-DDS` on `ghz` and `wstate` circuits. The performance variations among MPS implementations underscore the importance of optimizing emulation methods for diverse simulation tasks. Additionally, we observe that two circuit classes (`realamp` and `su2rand`) exhibit remarkably similar performance for each emulator, suggesting that these circuits may represent a single test case in terms of computational complexity. Similarly, the circuit classes `qftentangled` and `qft`, as well as `qpeexact` and `qpeinexact`, respectively show very similar performance for each MPS emulator, suggesting that the two circuits exhibit comparable entanglement structures.

## 4 COMPETITIVE ANALYSIS

Having conducted a comprehensive study of quantum emulators across diverse circuit classes, a natural question arises: which emulator demonstrates the best overall performance? Our study, comprising 759 benchmark instances across seven quantum emulators, 13 scalable circuit classes and qubit counts ranging from 4 to 1,024, provides a unique opportunity to address this question.

In general, we may observe that emulator A outperforms emulator B on a subset of tasks, while emulator B surpasses emulator C on another subset. However, such comparisons do not necessarily establish a clear hierarchy among emulators A, B, and C. To systematically compare emulators and to rank their relative "universality" on a wide range of quantum circuits, we propose a competitive evaluation framework based on the Elo rating system. Originally developed for chess rankings, the Elo system has been successfully adapted to compare performance in various domains, including optimization algorithms, programming competitions and large language models [38, 39]. The Elo system provides a robust method to measure the relative strength of emulators and to dynamically update ratings over time as new benchmark sets or emulators are introduced.

To compute the Elo ratings, we conduct pairwise competitions across all 13 benchmark algorithms. For each benchmark, we consider each emulator's best performance. In a head-to-head matchup, emulator A "wins" over B if it successfully simulates a higher number of qubits with fidelity $\mathcal{F} \geq 0.99$ in the 300-second time limit; if both simulators reach the same qubit count, the one with the lower total run time time wins. Initial Elo scores are set to 1,200, with updates calculated via iterative competitions over all valid problem instances. To ensure statistical robustness, we compute mean Elo scores and standard deviations across 200,000 independent trials with randomized algorithm orderings (see Appendix G for details on the Elo update procedure).

The final Elo rankings given in Table 3 show that the four top performers are MPS emulators, with `MIMIQ` outperforming `Qiskit` and `QMatchaTea`. It is important to note that `MIMIQ-MPS` underwent improvements during the course of this project, particularly through the inclusion of circuit pre-optimization steps, which had a significant impact on performance for the `graphstate` and `qnn` benchmarks. To promote transparency and encourage further advancements, we are releasing these results along with the benchmarking framework as an open resource. Additionally, we will rerun the benchmarks on the latest emulator releases prior to finalizing this publication to ensure accuracy and fairness. For reference, a difference of 200 Elo points indicates that the higher-rated emulator is expected to outperform the lower-rated emulator approximately 75% of the time. In the context of chess, this difference typically corresponds to distinct skill categories.

| emulator | Elo Average | Std |
|---|---|---|
| MIMIQ-MPS | 1,529 | 16 |
| Qiskit-MPS | 1,435 | 40 |
| QMatchaTea-MPS | 1,241 | 34 |
| Quimb-MPS | 1,132 | 51 |
| Pyqrack | 1,030 | 51 |
| MQT-DDS | 1,026 | 55 |
| Quimb-TN | 1,005 | 37 |

Table 3. **Final Elo Rankings.** Average emulator Elo scores (higher values indicate better performance) and standard deviations computed over 13 games and 200,000 randomized sequences.

The one-to-one win rate comparisons shown in Figure 3 demonstrate the internal consistency of the Elo ranking system. Specifically, if emulator $A$ has a higher Elo score than emulator $B$, the probability of $A$ outperforming $B$ is consistently above 50%. While the results are dependent on the specific selection of algorithms in this study, the benchmark set includes sufficient diversity in circuit types and difficulty levels to provide a meaningful evaluation of overall emulator performance.

Looking ahead, we envision that this ranking approach could be extended to larger and more comprehensive sets of circuits. Future studies could further refine the framework by categorizing circuits into specific types – such as variational algorithms, error correction protocols, or quantum machine learning tasks – to enable a more fine-grained and application-specific assessment of emulator performance for real-world use cases.

## 5 CONCLUSIONS

In this work, we conducted a comprehensive benchmarking study of seven utility-scale quantum emulators utilizing Tensor Network (TN), Matrix Product State (MPS), compressed ket states and decision diagram methodologies. Leveraging a suite of 13 scalable benchmark circuits from the MQTBench library with qubit counts ranging from 4 to 1,024, we systematically evaluated emulator performance in terms of run time and scalability in the effectively exact regime. Our findings highlight the superior scalability and efficiency of MPS-based methods across a broad range of circuit types, successfully simulating 8 out of 13 at the maximum scale of 1,024 qubits and 12 out of 13 benchmarks at 100 qubits, with a deep and wide random circuit as the only exception. The difficulty of simulating random quantum circuits on classical computers is well established, e.g., in the context of quantum supremacy experiments. However, our benchmarking results demonstrate that a broad spectrum of structured circuits, which have practical applications, exhibit polynomial run time scaling and can be efficiently simulated classically even beyond 100 qubits. This is particularly remarkable given that the tested emulators are universal, i.e. they have not been optimized for any particular circuit class. This also helps refine assumptions about quantum hardware requirements for quantum utility and informs future benchmarking studies by highlighting that circuit structure can be more critical than qubit count when evaluating quantum algorithms against state-of-the-art classical methods. It also challenges conventional assumptions that MPS techniques are limited to approximate simulations or low-connectivities, and highlights the need for more challenging benchmark suites for quantum computers, capable of pushing the limits of classical simulation.

A key contribution of this study is the development of a robust benchmarking methodology designed to ensure fair and reproducible comparisons across diverse quantum emulators. This methodology incorporates three critical components:

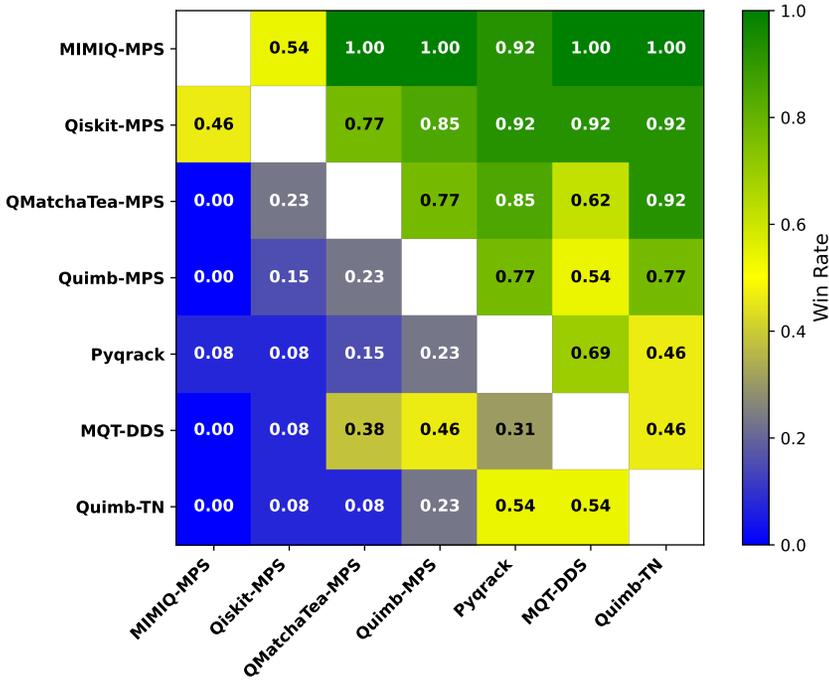

Fig. 3. **Heatmap visualization of the win rates between emulator pairs**. The matrix shows the probability of the emulators on the vertical axis outperforming the emulator on the horizontal axis across all benchmark tasks. Color intensity represents the win rate magnitude, from dark blue (0) to green (1), with corresponding numerical values displayed in each cell. Emulators are ordered from highest Elo rating (MIMIQ-MPS) at the top/left to lowest (Quimb-TN) at the bottom/right.

- Scalable benchmark sets to assess emulator performance over a wide range of problem sizes and difficulty levels.
- Automated hyperparameter tuning, leveraging the CMA-ES algorithm, to ensure each emulator achieves peak performance without possible biases associated with manual parameter selection.
- Mirror-circuit fidelity validation to independently test the accuracy of quantum emulations.

Our results reveal significant variations in performance, even among emulators based on similar underlying methodologies. No single emulator consistently outperformed others across all circuit types, underscoring room for further innovation in quantum emulation techniques. Therefore, we consider this study as a snapshot of the current state-of-the-art, which will continue to improve. We find that circuit pre-optimization and emulator specific parameters can have a large effect on performance which in some cases can be as important as the emulation method itself. This highlights an opportunity for AI-driven adaptive tuning strategies to dynamically adjust parameters based on circuit structure and qubit count, which could significantly improve performance and simplify user experience. In addition, incorporating circuit pre-conditioning steps, e.g. reducing the total number of gates, identifying common circuit primitives or optimizing gate sequences to limit memory usage or entanglement growth, can also improve performance significantly on certain circuits. While here we focused on CPU based implementations, we expect additional performance gains may be possible through hardware acceleration and GPU-based parallelization.

The benchmark protocols developed here are highly adaptable and can be readily extended to diverse hardware configurations, including GPU-based systems, supercomputers, and even genuine quantum computers. Future studies should focus on further diversifying problems sets, diversifying algorithms and hardware. Our run time metric can also be readily adapted for approximate regime simulations, for instance, to measure the total time required to obtain a correct sample after the mirror circuit, rather than the time to generate a fixed number of samples. To engage the community, we have published our source code in the open-access GitHub repository[4].

## 6 ACKNOWLEDGMENTS

This work has benefited from a state grant managed by the French National Research Agency under the Investments of the Future Pro- gram with the reference ANR-21-ESRE-0032 "aQCess - Atomic Quantum Computing as a Service", the Horizon Europe programme HORIZON-CL4-2021-DIGITAL-EMERGING-01-30 via the project "EuRyQa - European infrastructure for Rydberg Quantum Computing" grant agreement number 10107014 and support from the Institut Universitare de France (IUF).

## 7 COMPETING INTERESTS

GM and SW are co-founders and shareholders of QPerfect.

---

[4]https://github.com/qperfect-io/feniqs_lite

## A SELECTED EMULATORS

The selected emulators and their underlying methodologies are as follows:

(1) `Qiskit` v2.0.0 (`Qiskit-Aer` v0.17.0) – open-source MPS emulator [40]
(2) `Quimb` v1.8.4 – open-source TN emulator [41]
(3) `Quimb-MPS` v1.8.4 – open-source MPS emulator [41]
(4) `QMatchaTea-MPS` v1.1.3 – open-source MPS emulator
(5) `MQT-DDS` v1.24 – open-source Decision Diagram emulator [42]
(6) `Pyqrack` v1.35.6 – open-source qubit and gate emulator based on factorized ket simulation [43]
(7) `MIMIQ-MPS` v0.17.4 – commercial MPS emulator by QPerfect.

While we selected these 7 emulators, we also considered the following emulators, which were excluded on the basis that to the best of our knowledge they were either no longer actively developed, unable to execute a universal gate set, limited in qubit numbers, or unable to reach the 100 qubit scale: TensorCircuit, TeNPy, PastaQ, QuantUnity, Tensorly, QTensor, QXTools, QFlex, Qibo (Quimb backend), TeD-Q, TnQVM, QuCircuit, JuliVQC, Tree Tensor Network (TTN), Pennylane (Quimb backend), Qrisp, QMatchaTea-TTN, Cirq (statevector only), Yao (statevector only), Qulacs (statevector only).

## B BENCHMARK SUITE

The benchmark suite consists of the following 13 circuit classes (from the MQTBench library [5]):

(1) Amplitude Estimation (ae[6]): A quantum algorithm used to estimate the amplitude of a given quantum state, crucial for applications in quantum finance [44].
(2) GHZ State (ghz): A simple circuit for generating a fully entangled Greenberger-Horne-Zeilinger state [45].
(3) W-State (wstate): An entangled state distinct from GHZ states, studied for its applications in quantum networks [46].
(4) Graph State (graphstate[7]): A quantum circuit to prepare a graph state by preparing all qubits in the "+" state and then applying a CZ gate for each edge of the graph. Graph State is a Clifford circuit used in quantum error correction [47].
(5) Quantum Fourier Transform (qft): it is the quantum equivalent of the discrete Fourier transform and is a very important building block in many quantum algorithms. [48].
(6) Entangled QFT (qftentangled): Quantum Fourier Transform applied to an entangled GHZ state.
(7) Quantum Phase Estimation Exact (qpeexact): Estimates the phase of a quantum operation and is a very important building block in many quantum algorithms. In the exact case, the applied phase is exactly representable by the number of qubits. [49].
(8) Quantum Phase Estimation Inexact (qpeinexact): Estimates the phase of a quantum operation and is a very important building block in many quantum algorithms. In the inexact case, the applied phase is not exactly representable by the number of qubits.
(9) VQE Ansatz with Real Amplitudes (realamp[8]): Realizes a VQE ansatz circuit with randomly initialized values [50]. This is a heuristic trial wave function used as Ansatz in chemistry applications or classification circuits in machine learning. The circuit consists of alternating layers of Y rotations and CX entangling gates. It only produces real valued statevector amplitudes.
(10) VQE Ansatz with SU2 Random Parameters (su2rand[9]): Realizes a VQE ansatz circuit consisting of layers of single qubit operations spanned by SU(2) with random parameters and CX entangling gates. It is a heuristic pattern that can be used to prepare trial wave functions for variational quantum algorithms or classification circuits for machine learning.
(11) Quantum Walk V-Chain (qwalk): Realizes a quantum random walk using a v-chain as ancillae qubits [51].
(12) Random Circuit (random): Realizes a random circuit which is twice as deep as wide. It considers random quantum gates with up to four qubits [17].
(13) Quantum Neural Network (qnn[10]): Quantum Neural Network [52] with a ZZ FeatureMap and a RealAmplitudes ansatz.

One additional VQE Ansatz twolocal was excluded on the basis that it produces identical circuits to realamp. Other benchmark circuits from the MQTBench library were excluded on the basis that they were unable to scale beyond 100 qubits.

---

[5]https://www.cda.cit.tum.de/mqtbench/benchmark_description
[6]https://qiskit.org/documentation/finance/tutorials/00_amplitude_estimation.html
[7]https://docs.quantum.ibm.com/api/qiskit/qiskit.circuit.library
[8]https://docs.quantum.ibm.com/api/qiskit/qiskit.circuit.library.RealAmplitudes
[9]https://docs.quantum.ibm.com/api/qiskit/qiskit.circuit.library.EfficientSU2
[10]https://qiskit.org/ecosystem/machine-learning/stubs/qiskit_machine_learning.neural_networks.EstimatorQNN.html

## C TUNING PARAMETERS FOR SELECTED SIMULATORS

Below is a summary of the tuning parameters and their possible values for each simulator.

**Qiskit-MPS**

Parameters:
[ `matrix_product_state_max_bond_dimension` /
`matrix_product_state_truncation_threshold` /
`mps_sample_measure_algorithm` ],
[ `optimization_level` / `mps_lapack` ]

- Convergence parameters:
  - `matrix_product_state_max_bond_dimension`: Maximum bond dimension allowed (e.g., 4, 8, 16, 32, 64, 128, 256, 512, 1024, 2048, 3072).
  - `matrix_product_state_truncation_threshold`: MPS truncation threshold (from $1 \times 10^{-1}$ to $1 \times 10^{-10}$).
  - `mps_sample_measure_algorithm`: Measurement simulation algorithm ("P" for MPS probabilities; "AM" for apply measure).
- Method parameters:
  - `optimization_level`: Optimization level in circuit compilation (1, 2, or 3).
  - `mps_lapack`: Option for using `OpenBLAS/Lapack` interface for SVD ("F" for False, "T" for True).

See the Qiskit documentation.

**MIMIQ-MPS**

Parameters:
[ `bond_dimension` / `entdim` / `scut` ],
[ `meth` / `fuse` / `perm` ]

- Convergence parameters:
  - `bond_dimension`: Maximum bond dimension allowed (e.g., 4, 8, 16, 32, 64, 128, 256, 512, 1024, 2048, 3072).
  - `entdim`: Maximum entanglement dimension allowed (4, 8, or 16).
  - `scut`: A fixed threshold for discarding small singular values during truncation (from $1 \times 10^{-1}$ to $1 \times 10^{-10}$).
- Method parameters:
  - `meth`: MPS implementation method (e.g., vmpoa, dmpo).
    `dmpo` – Direct Matrix Product Operator (MPO): Gates are converted and compressed into MPO form. Each MPO is then directly applied with tensor contractions to the quantum state in MPS form.
    `vmpoa` – Variational MPO type A: Gates are converted and compressed into MPO form. Each MPO is then applied to the quantum state in MPS form by performing tensor network contractions and by variationally optimizing the resulting state in order to maximize its fidelity.
  - `fuse`: Fuse together successive gates on the same qubits (up to 2 qubit gates) before compression into matrix product operators. ("F" – False, "T" – True)
  - `perm`: Allows optimization of qubit ordering when mapping the algorithm to the 1-dimensional MPS topology to minimize entanglement in MPS form ("F" – False, "T" – True).

Refer to the MIMIQ documentation.

**QMatchaTea-MPS**

Parameters:
 [ `max_bond_dimension` / `cut_ratio` ],
 [ `optimization_level` / `tensor_compilator_enable` / `linearize_enable` ]

- Convergence parameters:
  - `max_bond_dimension`: Maximum bond dimension allowed (e.g., 4, 8, 16, 32, 64, 128, 256, 512, 1024, 2048, 3072).
  - `cut_ratio`: Contraction ratio threshold (from $1 \times 10^{-1}$ to $1 \times 10^{-10}$).
- Method parameters:
  - `optimization_level`: Optimization level (0, 1, or 2).
  - `tensor_compilator_enable`: Flag for using the tensor compiler ("F" – False, "T" – True).
  - `linearize_enable`: Flag indicating circuit linearization ("F" – False, "T" – True).

For more details, see the QMatchaTea documentation.

**Quimb-MPS**

Parameters:
 [ `bonddim` / `cutoff` ],
 [ `permute` / `gate_contract` ]

- Convergence parameters:
  - `bonddim`: Maximum bond dimension allowed.
  - `cutoff`: Truncation threshold.
- Method parameters:
  - `permute`: Flag for a permutation-based strategy that applies non-local two-qubit gate contractions using swap+split ("F" – False, "T" – True).
  - `gate_contract`: Mode for gate contraction; use AU for the auto-MPS approach or SS for forcing swap+split.

See the Quimb documentation.

**Quimb-TN**

Parameter:
 [ `contract` ]

- Method parameter:
  - `contract`: Flag to perform gate optimization ("F" – False, "T" – True).

See the Quimb documentation.

**Other Emulators**

The emulators `Pyqrack` and `MQT-DDS` do not offer adjustable tuning parameters.

## D CLASSIFICATION OF CIRCUIT COMPLEXITY BASED ON SIMULATION DIFFICULTY

| Algorithm | 4 | 8 | 16 | 24 | 32 | 64 | 128 | 256 | 512 | 1024 |
|---|---|---|---|---|---|---|---|---|---|---|
| ae | 4/2/1 * MIMIQ | 4/2/1 * Qiskit | 4/2/1 * Qiskit | 4/2/0 * MIMIQ | 4/1/0 * MIMIQ | 4/0/0 * MIMIQ | 4/0/0 * Qiskit | 3/0/0 * Qiskit | 2/0/0 * Qiskit | 2/0/0 * Qiskit |
| ghz | 4/2/1 * MIMIQ | 4/2/1 * MIMIQ | 4/2/1 * MIMIQ | 4/2/1 * MIMIQ | 4/2/1 * MIMIQ | 4/2/1 * MIMIQ | 4/1/1 * MIMIQ | 4/0/1 * MIMIQ | 4/0/1 * MIMIQ | 3/0/1 * MIMIQ |
| graphstate | 4/2/1 * MIMIQ | 4/2/1 * MIMIQ | 4/2/1 * Pyqrack | 4/2/1 * MIMIQ | 3/2/1 * Qiskit | 2/2/0 * MIMIQ | 1/1/0 * MIMIQ | 1/0/0 * MIMIQ | 1/0/0 * MIMIQ | 1/0/0 * MIMIQ |
| qft | 4/2/1 * MIMIQ | 4/2/1 * MIMIQ | 4/2/1 * MQT-DDS | 4/2/1 * MQT-DDS | 4/2/1 * MQT-DDS | 4/2/1 * MQT-DDS | 3/0/0 * Qiskit | 3/0/0 * Qiskit | 2/0/0 * Qiskit | 2/0/0 * Qiskit |
| qftentangled | 4/2/1 * MIMIQ | 4/2/1 * MIMIQ | 4/2/1 * MIMIQ | 4/1/0 * Qiskit | 4/0/0 * Qiskit | 4/0/0 * Qiskit | 3/0/0 * Qiskit | 3/0/0 * Qiskit | 2/0/0 * Qiskit | 2/0/0 * Qiskit |
| qnn | 4/2/1 * MIMIQ | 4/2/1 * MIMIQ | 3/2/1 * Qiskit | 3/1/0 * Qiskit | 3/0/0 * QMatchaTea | 2/0/0 * MIMIQ | 0/0/0/0 * None | 0/0/0/0 * None | 0/0/0/0 * None | 0/0/0/0 * None |
| qpeexact | 4/2/1 * MIMIQ | 4/2/1 * MIMIQ | 4/2/1 * MIMIQ | 4/2/1 * Qiskit | 4/1/0 * Qiskit | 4/0/0 * Qiskit | 3/0/0 * Qiskit | 2/0/0 * Qiskit | 2/0/0 * Qiskit | 2/0/0 * Qiskit |
| qpeinexact | 4/2/1 * MIMIQ | 4/2/1 * MIMIQ | 4/2/1 * MIMIQ | 4/2/0 * MIMIQ | 4/1/0 * MIMIQ | 4/0/0 * Qiskit | 3/0/0 * Qiskit | 2/0/0 * Qiskit | 2/0/0 * Qiskit | 2/0/0 * Qiskit |
| qwalk | 4/2/1 * MIMIQ | 4/2/1 * Qiskit | 4/1/1 * MQT-DDS | 4/0/1 * MQT-DDS | 4/0/1 * MQT-DDS | 3/0/1 * MQT-DDS | 2/0/1 * MQT-DDS | 1/0/0 * Qiskit | 0/0/0 * None | 0/0/0 * None |
| random | 4/2/1 * Pyqrack | 4/2/1 * Qiskit | 3/1/1 * Pyqrack | 0/1/0 * Pyqrack | 0/0/0/0 * None | 0/0/0/0 * None | 0/0/0/0 * None | 0/0/0/0 * None | 0/0/0/0 * None | 0/0/0/0 * None |
| realamp | 4/2/1 * MIMIQ | 4/2/1 * MIMIQ | 4/2/1 * MIMIQ | 4/1/0 * MIMIQ | 4/0/0 * MIMIQ | 3/0/0 * MIMIQ | 1/0/0 * MIMIQ | 1/0/0 * MIMIQ | 0/0/0 * None | 0/0/0 * None |
| su2rand | 4/2/1 * MIMIQ | 4/2/1 * MIMIQ | 4/2/1 * MIMIQ | 4/1/0 * MIMIQ | 4/0/0 * MIMIQ | 3/0/0 * MIMIQ | 1/0/0 * MIMIQ | 1/0/0 * MIMIQ | 0/0/0 * None | 0/0/0 * None |
| wstate | 4/2/1 * MIMIQ | 4/2/1 * MIMIQ | 4/2/1 * MIMIQ | 4/2/1 * MIMIQ | 4/2/1 * MIMIQ | 4/1/1 * MIMIQ | 4/0/1 * MIMIQ | 4/0/1 * MIMIQ | 4/0/1 * MIMIQ | 3/0/1 * MIMIQ |

Qubits

Fig. 4. **Benchmark complexity classification.** The matrix shows a complexity-based classification of the 13 quantum algorithms in the benchmark suite (vertical axis), for different qubit counts (horizontal axis) from 4 to 1024. Each cell displays the number of emulators from three categories (MPS/TN/DDS) that solved the instance within 300 seconds at a fidelity of at least 0.99, along with the best-performing emulator (labeled "*"). The cell color indicates empirical difficulty classes, as defined in the main text: easy (green), medium (orange), hard (red), and very hard (dark red). The simulator names (Qiskit, MIMIQ, and QMatchaTea) are shown without specifying their simulation type (i.e., MPS) to enhance the readability.

Recognizing some correlation between different circuit and tuning parameters, we attempt to classify benchmark circuits based on their empirical complexity. Figure 4 classifies each benchmark instance (i.e., a given algorithm at a particular qubit count) according to the number of emulators that can successfully solve it within a 300-second limit and with a fidelity of at least 0.99. For each problem instance, we calculate the "solved rate" as the ratio of emulators that meet these criteria to the total number of emulators evaluated. Instances where 60% or more of the emulators succeed are classified as "Easy," those with 30–60% as "Medium," 10–30% as "Hard," and less than 10% as "Very Hard." The color scale (green to red) reflects these empirical difficulty levels. Within each cell, we list the number of emulators from three categories—MPS, TN, DDS—that successfully solve the task, and the "*" field identifies the emulator achieving the shortest run time. In the cells of Figure 4, simulator names such as Qiskit, MIMIQ, and QMatchaTea are shown without specifying their simulation type (i.e., MPS) to enhance the readability. This visualization highlights which benchmarks remain tractable at high qubit counts and which become intractable at lower thresholds. Overall, this heatmap-like figure offers a sort of visual guide for selecting emulators based on task complexity.

# E  DETAILED SCALABILITY RESULTS

Table 4. **Emulator scalability**. The table shows the largest number of qubits each emulator could successfully process within the 300-second runtime limit while maintaining a mirror fidelity of at least 0.99. Data is organized by algorithm (rows) and emulator implementation (columns), with higher values indicating better scalability. Note: for `qwalk`, we count the total number of qubits, including ancilla qubits.

| alg | Qiskit MPS | MIMIQ MPS | QMatchaTea MPS | Quimb MPS | Quimb TN | Pyqrack | MQT DDS |
|---|---|---|---|---|---|---|---|
| ae | 1024 | 1024 | 256 | 128 | 40 | 28 | 16 |
| ghz | 1024 | 1024 | 1024 | 640 | 128 | 120 | 1024 |
| graphstate | 96 | 1024 | 44 | 28 | 128 | 72 | 60 |
| qft | 1024 | 1024 | 256 | 112 | 36 | 30 | 80 |
| qftentangled | 1024 | 1024 | 256 | 112 | 18 | 28 | 16 |
| qnn | 52 | 120 | 100 | 14 | 18 | 26 | 18 |
| qpeexact | 1024 | 1024 | 224 | 112 | 60 | 30 | 28 |
| qpeinexact | 1024 | 1024 | 224 | 112 | 40 | 30 | 16 |
| qwalk | 349 | 221 | 61 | 85 | 13 | 25 | 157 |
| random | 16 | 18 | 16 | 12 | 14 | 26 | 16 |
| realamp | 32 | 352 | 100 | 100 | 18 | 28 | 16 |
| su2rand | 32 | 352 | 100 | 100 | 18 | 28 | 16 |
| wstate | 1024 | 1024 | 1024 | 640 | 120 | 32 | 1024 |

# F  EARLY TERMINATION CONDITIONS AND CRASH REPORTS

This section lists problem instances that failed before reaching the 300-second timeout (or when the fidelity value is below 0.99). The number in parentheses indicates the number of qubits at which the failure condition occurred.

- **ae**
  - Quimb-MPS (144): *(i) for 144 qubits – the simulation finished with the fidelity below 0.99; (ii) from 160 qubits – by timeout.*
  - Pyqrack (30): *(i) for 30–32 qubits - by timeout; (ii) from 36 qubits - the simulation halts automatically when an internal stop condition is met – the fidelity falls below the acceptable threshold.*
- **ghz**
  - Quimb-TN (144): *from quimb/tensor/circuit.py, sample_bitstring_from_prob_ndarray: ValueError: probabilities contain NaN.*
  - Pyqrack (128): *Double free or corruption (out).*
- **qft**
  - QMatchaTea (288): *from qmatchatea/preprocessing.py, TranspilerError: Swap mapper failed: layer 2293, sublayer 1.*
  - MQT-DDS (88): *Numerical instabilities led to a 0-vector! Abort simulation!*
  - Pyqrack (32): *(i) for 32 qubits - by timeout; (ii) from 36 qubits - the simulation halts automatically when an internal stop condition is met – the fidelity falls below the acceptable threshold.*
- **qftentangled**
  - QMatchaTea (288): *from qmatchatea/preprocessing.py, TranspilerError: Swap mapper failed: layer 3043, sublayer 5*
  - Pyqrack (30): *(i) for 30–32 qubits - by timeout; (ii) for 36–60 qubits - the simulation halts automatically when an internal stop condition is met – the fidelity falls below* *the acceptable threshold; (iii) for 64 qubits - QEngineCPU::Apply2x2 offset1 and offset2 parameters must be within allocated qubit bounds! (iv) from 72 qubits - Cannot instantiate a register with greater capacity than native types on emulating system.*

- **graphstate**
  - Qiskit (100): *qiskit.exceptions.QiskitError: 'ERROR: [Experiment 0] Error: Wrong SVD calculations: A != USV* , ERROR: Error: Wrong SVD calculations: A != USV*'*
  - QMatchaTea (48): *by timeout, but with warning – from qtealeaves/tensors/tensor.py: gesdd SVD decomposition failed. Resorting to gesvd.*
  - MQT-DDS (64): *(i) for 64–160 qubits – by timeout; (ii) from 176 qubits - RuntimeError: Numerical instabilities led to a 0-vector! Abort simulation!*
  - Pyqrack (80): *the simulation halts automatically when an internal stop condition is met – the fidelity falls below the acceptable threshold.*
  - Quimb-TN (144): *from quimb/tensor/circuit.py, sample_bitstring_from_prob_ndarray: ValueError: probabilities contain NaN*
- **qnn**
  - Pyqrack (28): *(i) for 28–32 qubits - by timeout; (ii) from 36 qubits - the simulation halts automatically when an internal stop condition is met – the fidelity falls below the acceptable threshold.*
- **qpeexact**
  - Pyqrack (32): *(i) for 32–36 qubits - by timeout; (ii) from 40 qubits - the simulation halts automatically when an internal stop condition is met – the fidelity falls below the acceptable threshold.*
- **qpeinexact**
  - Pyqrack (32): *(i) for 32–36 qubits - by timeout; (ii) from 40 qubits - the simulation halts automatically when an internal stop condition is met – the fidelity falls below the acceptable threshold.*
- **qwalk**
  - MQT-DDS (173): *(i) for 173 – by timeout; (ii) from 189 qubits – RuntimeError: Numerical instabilities led to a 0-vector! Abort simulation!*
  - Pyqrack (27): *(i) for 27–69 qubits - by timeout; (ii) from 77 qubits - the simulation halts automatically when an internal stop condition is met – the fidelity falls below the acceptable threshold.*
- **realamp**
  - Pyqrack (30): *(i) for 30–32 qubits - by timeout; (ii) from 36 qubits - the simulation halts automatically when an internal stop condition is met – the fidelity falls below the acceptable threshold.*
- **su2rand**
  - Pyqrack (30): *(i) for 30–32 qubits - by timeout; (ii) from 36 qubits - the simulation halts automatically when an internal stop condition is met – the fidelity falls below the acceptable threshold.*
- **wstate**
  - Pyqrack (36): *from 36 qubits - the simulation halts automatically when an internal stop condition is met – the fidelity falls below the acceptable threshold.*
- **random**
  - Pyqrack (28): *(i) for 28–32 qubits - by timeout; (ii) from 36 qubits - the simulation halts automatically when an internal stop condition is met – the fidelity falls below the acceptable threshold.*

– MIMIQ (20): *(i) for 20 qubits – the simulation finished with the fidelity below 0.99; (ii) from 24 qubits – by timeout.*

## G ELO-BASED RANKING METRIC

For two simulators $A$ and $B$ with current Elo ratings $R_A$ and $R_B$, the updated rating $R'_A$ for emulator $A$ post-comparison is computed as:

$$R'_A = R_A + K(S_A - E_A), \qquad (2)$$

where:
- $S_A$ represents the actual outcome: $S_A = 1$ if emulator $A$ wins (achieves a better rank), and $S_A = 0$ if emulator $A$ loses.
- $K$ is the maximum rating adjustment factor, set to 32 in this study to balance stability and adaptability.
- $E_A$ is the expected probability of winning, calculated as:

$$E_A = \frac{1}{1 + 10^{(R_B - R_A)/400}}. \qquad (3)$$

## H RUN TIME PLOTS

For providing detailed insights into the performance of the emulators, we present the individual run time plots with respect to the number of qubits for each benchmark task. Each figure shows the run time (in seconds, on a logarithmic scale) as a function of the number of qubits for a given algorithm. The post-processing scripts are available in the GitHub repository[11].

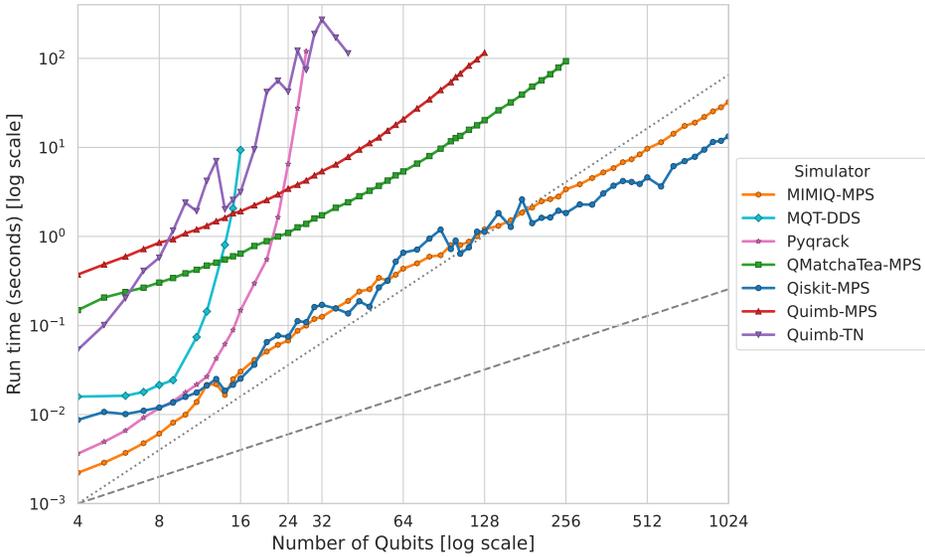

Fig. 5. **Amplitude Estimation (ae) benchmark**. The log-log plot displays number of qubits from 4 to 1024 (vertical axis) against computation time in seconds (horizontal axis) for the seven tested simulators: MQT-DDS (light blue diamonds), MIMIQ-MPS (orange circles), Pyqrack (pink stars), QMatchaTea-MPS (green squares), Qiskit-MPS (dark blue circles), Quimb-MPS (red triangles), and Quimb-TN (purple downward-pointing triangles).

---
[11]https://github.com/qperfect-io/feniqs_lite

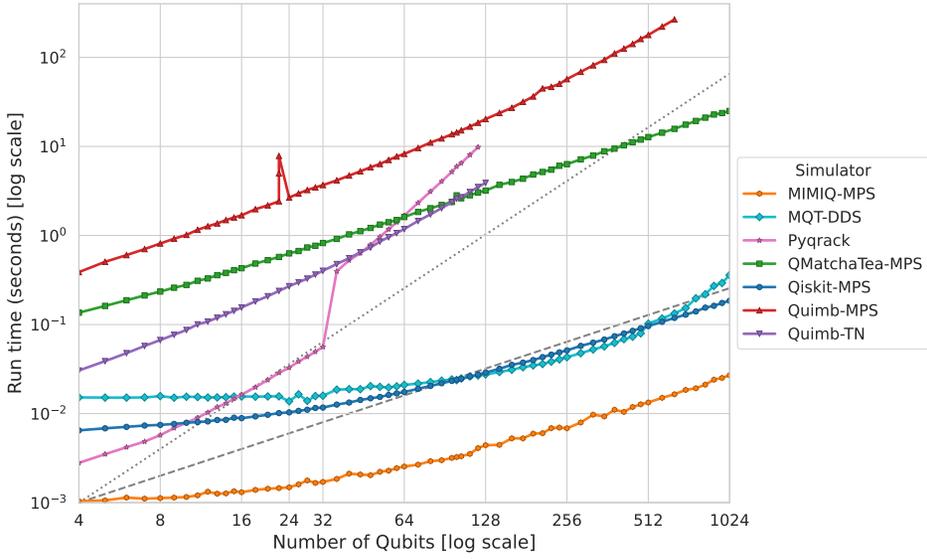

Fig. 6. **GHZ State (ghz) benchmark**. The log-log plot displays number of qubits from 4 to 1024 (vertical axis) against computation time in seconds (horizontal axis) for the seven tested simulators: MQT-DDS (light blue diamonds), MIMIQ-MPS (orange circles), Pyqrack (pink stars), QMatchaTea-MPS (green squares), Qiskit-MPS (dark blue circles), Quimb-MPS (red triangles), and Quimb-TN (purple downward-pointing triangles).

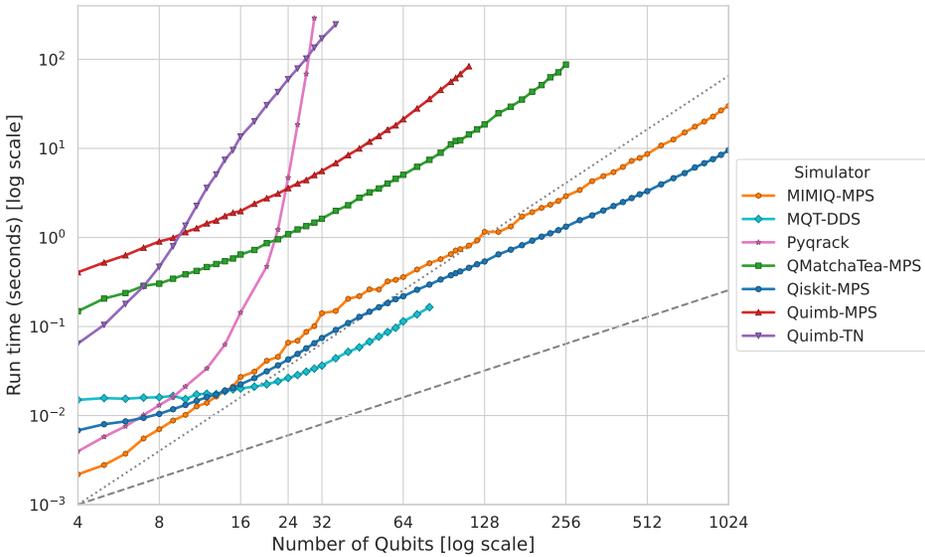

Fig. 7. **Quantum Fourier Transform (qft) benchmark**. The log-log plot displays number of qubits from 4 to 1024 (vertical axis) against computation time in seconds (horizontal axis) for the seven tested simulators: MQT-DDS (light blue diamonds), MIMIQ-MPS (orange circles), Pyqrack (pink stars), QMatchaTea-MPS (green squares), Qiskit-MPS (dark blue circles), Quimb-MPS (red triangles), and Quimb-TN (purple downward-pointing triangles).

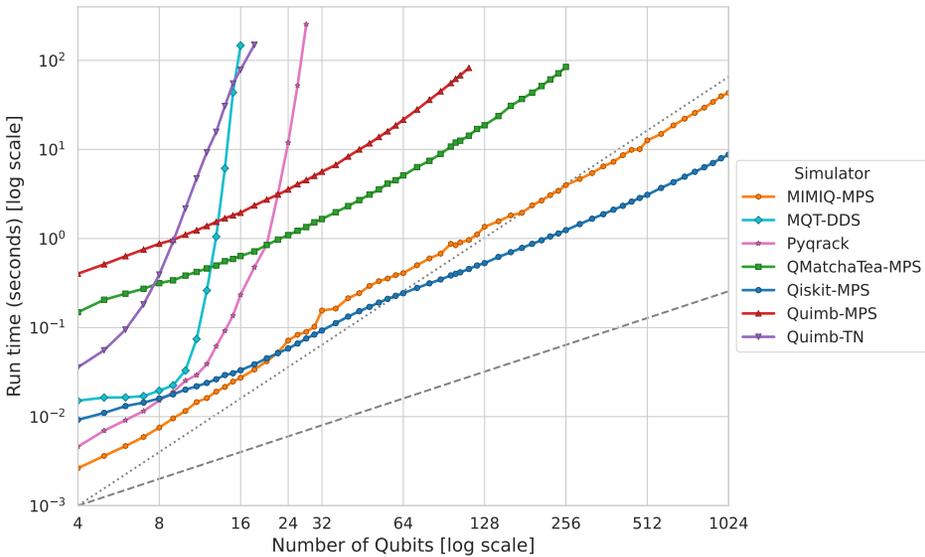

Fig. 8. **Entangled Quantum Fourier Transform (qftentangled) benchmark**. The log-log plot displays number of qubits from 4 to 1024 (vertical axis) against computation time in seconds (horizontal axis) for the seven tested simulators: MQT-DDS (light blue diamonds), MIMIQ-MPS (orange circles), Pyqrack (pink stars), QMatchaTea-MPS (green squares), Qiskit-MPS (dark blue circles), Quimb-MPS (red triangles), and Quimb TN (purple downward-pointing triangles).

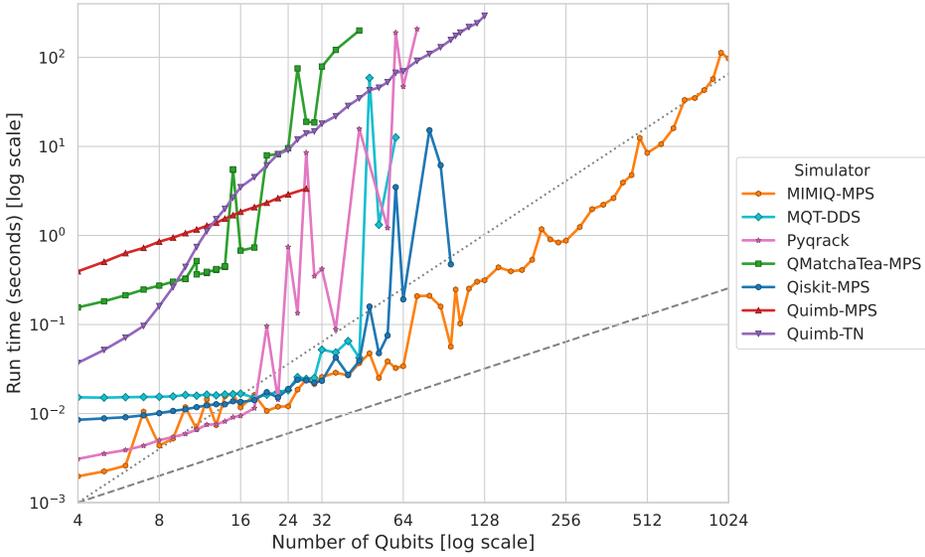

Fig. 9. **Graph State (graphstate) benchmark**. The log-log plot displays number of qubits from 4 to 1024 (vertical axis) against computation time in seconds (horizontal axis) for the seven tested simulators: `MQT-DDS` (light blue diamonds), `MIMIQ-MPS` (orange circles), `Pyqrack` (pink stars), `QMatchaTea-MPS` (green squares), `Qiskit-MPS` (dark blue circles), `Quimb-MPS` (red triangles), and `Quimb-TN` (purple downward-pointing triangles). The significant variability observed in the `graphstate` benchmark stems from its circuit design, which incorporates randomly chosen long-range gates that vary between instances.

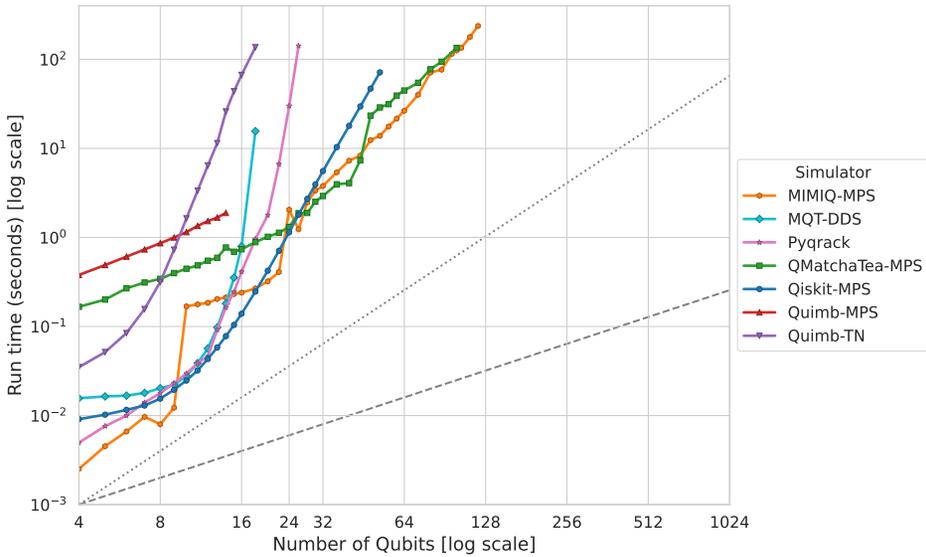

Fig. 10. **Quantum Neural Network (qnn) benchmark**. The log-log plot displays number of qubits from 4 to 1024 (vertical axis) against computation time in seconds (horizontal axis) for the seven tested simulators: MQT-DDS (light blue diamonds), MIMIQ-MPS (orange circles), Pyqrack (pink stars), QMatchaTea-MPS (green squares), Qiskit-MPS (dark blue circles), Quimb-MPS (red triangles), and Quimb-TN (purple downward-pointing triangles).

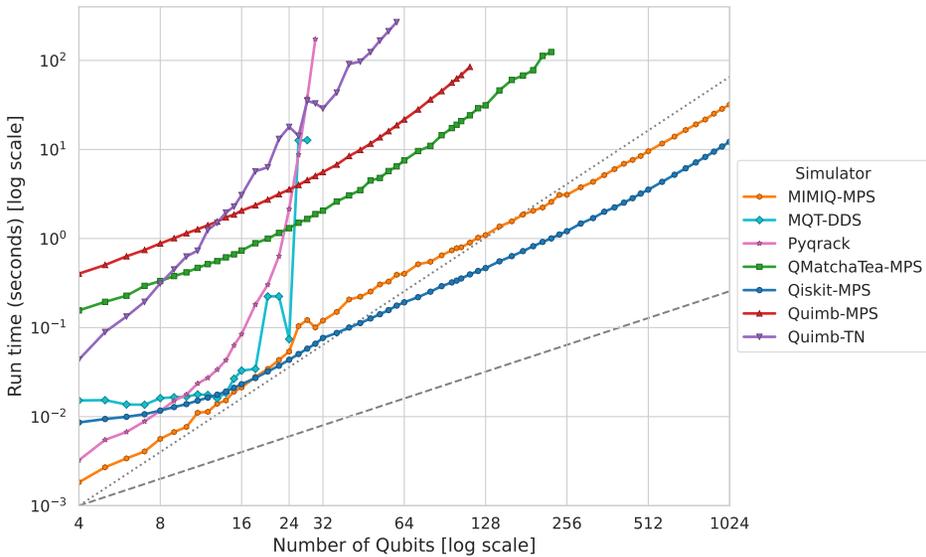

Fig. 11. **Exact Quantum Phase Estimation (qpeexact) benchmark**. The log-log plot displays number of qubits from 4 to 1024 (vertical axis) against computation time in seconds (horizontal axis) for the seven tested simulators: MQT-DDS (light blue diamonds), MIMIQ-MPS (orange circles), Pyqrack (pink stars), QMatchaTea-MPS (green squares), Qiskit-MPS (dark blue circles), Quimb-MPS (red triangles), and Quimb-TN (purple downward-pointing triangles).

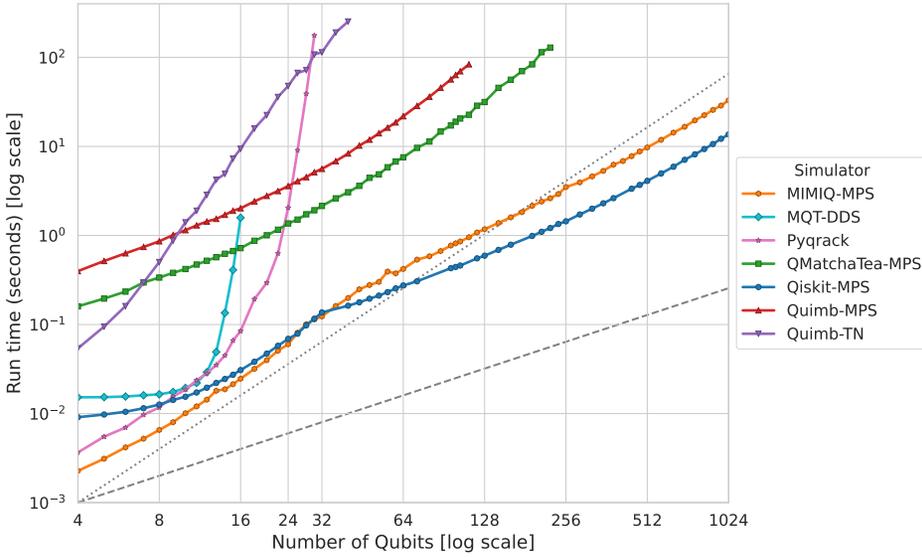

Fig. 12. **Inexact Quantum Phase Estimation (qpeinexact) benchmark**. The log-log plot displays number of qubits from 4 to 1024 (vertical axis) against computation time in seconds (horizontal axis) for the seven tested simulators: MQT-DDS (light blue diamonds), MIMIQ-MPS (orange circles), Pyqrack (pink stars), QMatchaTea-MPS (green squares), Qiskit-MPS (dark blue circles), Quimb-MPS (red triangles), and Quimb-TN (purple downward-pointing triangles).

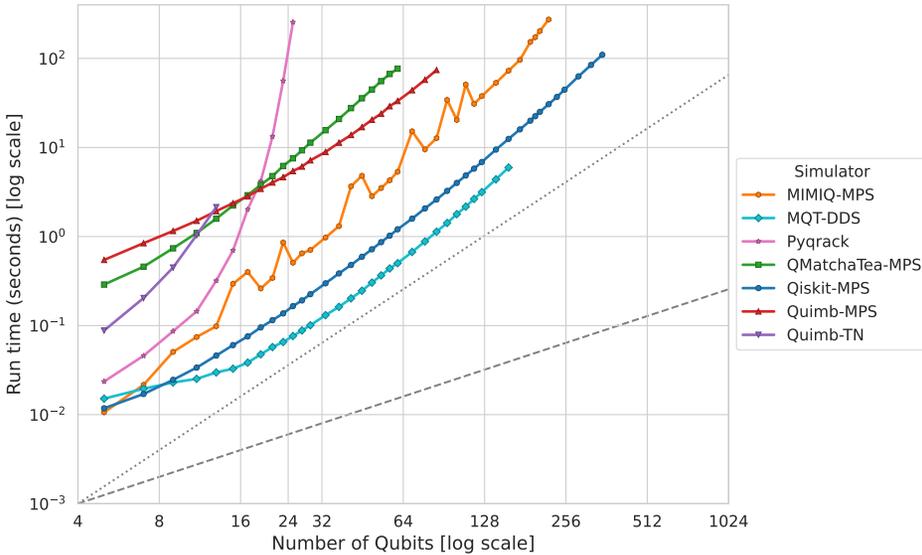

Fig. 13. **Quantum Walk V-Chain (qwalk) benchmark**. The log-log plot displays number of qubits from 4 to 1024 (vertical axis) against computation time in seconds (horizontal axis) for the seven tested simulators: MQT-DDS (light blue diamonds), MIMIQ-MPS (orange circles), Pyqrack (pink stars), QMatchaTea-MPS (green squares), Qiskit-MPS (dark blue circles), Quimb-MPS (red triangles), and Quimb-TN (purple downward-pointing triangles).

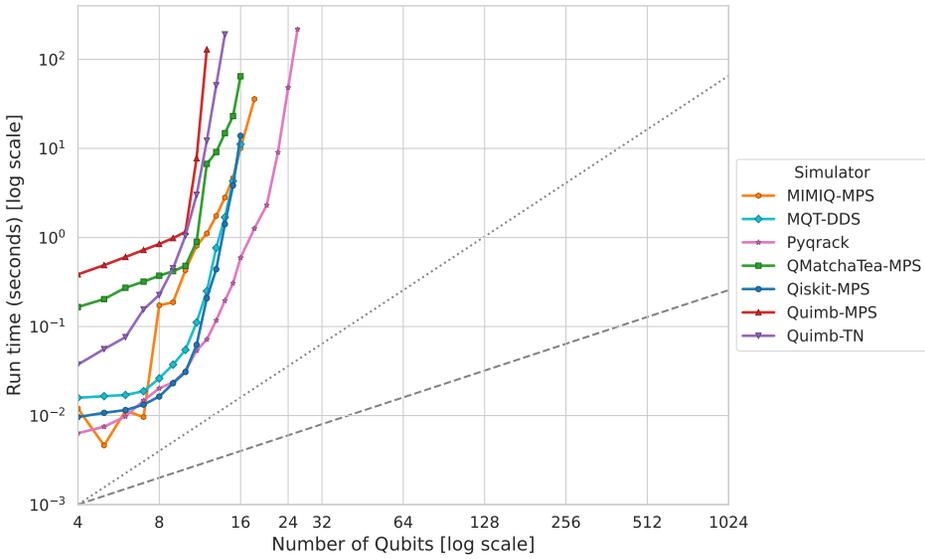

Fig. 14. **Random Circuit (random) benchmark**. The log-log plot displays number of qubits from 4 to 1024 (vertical axis) against computation time in seconds (horizontal axis) for the seven tested simulators: MQT-DDS (light blue diamonds), MIMIQ-MPS (orange circles), Pyqrack (pink stars), QMatchaTea-MPS (green squares), Qiskit-MPS (dark blue circles), Quimb-MPS (red triangles), and Quimb-TN (purple downward-pointing triangles).

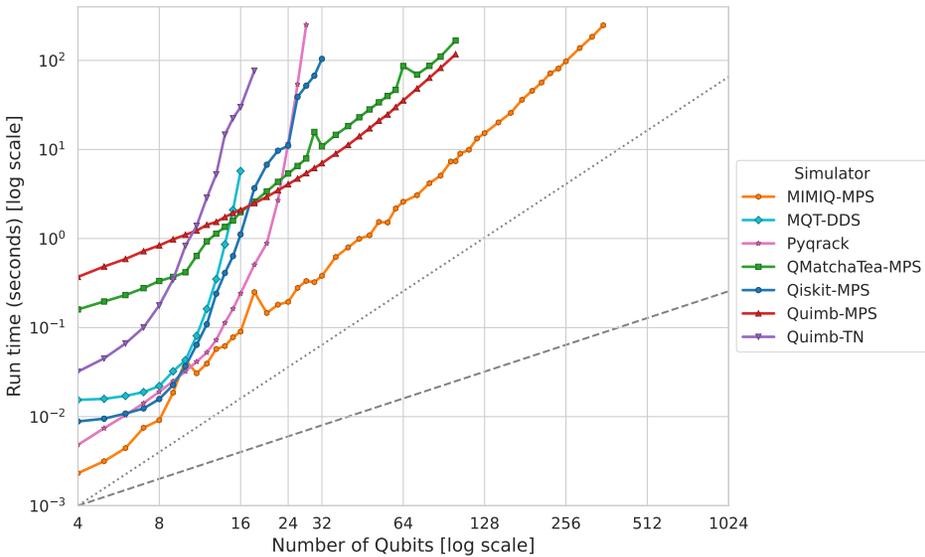

Fig. 15. **VQE Ansatz with Real Amplitudes (realamp) benchmark**. The log-log plot displays number of qubits from 4 to 1024 (vertical axis) against computation time in seconds (horizontal axis) for the seven tested simulators: MQT-DDS (light blue diamonds), MIMIQ-MPS (orange circles), Pyqrack (pink stars), QMatchaTea-MPS (green squares), Qiskit-MPS (dark blue circles), Quimb-MPS (red triangles), and Quimb-TN (purple downward-pointing triangles).

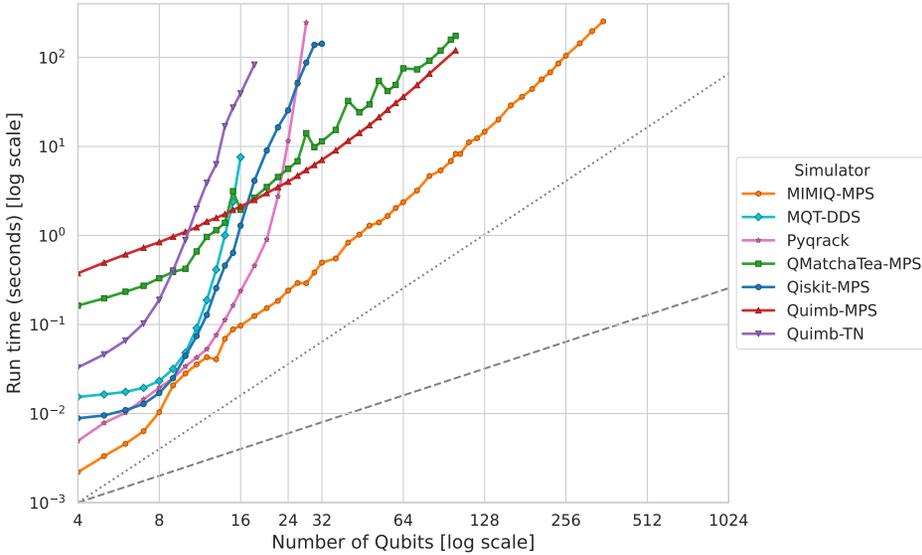

Fig. 16. **VQE Ansatz with SU2 Random Parameters (su2rand) benchmark**. The log-log plot displays number of qubits from 4 to 1024 (vertical axis) against computation time in seconds (horizontal axis) for the seven tested simulators: MQT-DDS (light blue diamonds), MIMIQ-MPS (orange circles), Pyqrack (pink stars), QMatchaTea-MPS (green squares), Qiskit-MPS (dark blue circles), Quimb-MPS (red triangles), and Quimb-TN (purple downward-pointing triangles).

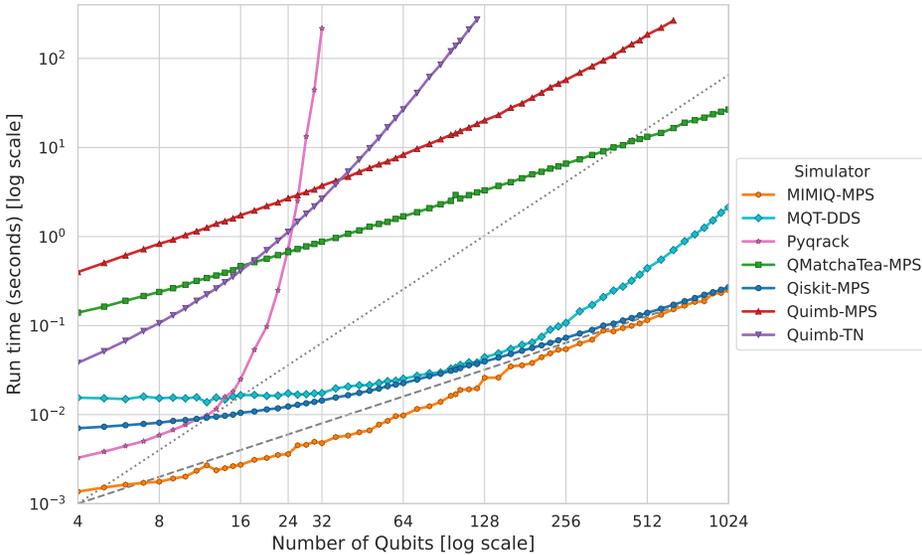

Fig. 17. **W-State (wstate) benchmark**. The log-log plot displays number of qubits from 4 to 1024 (vertical axis) against computation time in seconds (horizontal axis) for the seven tested simulators: MQT-DDS (light blue diamonds), MIMIQ-MPS (orange circles), Pyqrack (pink stars), QMatchaTea-MPS (green squares), Qiskit-MPS (dark blue circles), Quimb-MPS (red triangles), and Quimb-TN (purple downward-pointing triangles).